%% file: main.tex
\pdfoutput=1
\documentclass[usenatbib]{mnras}
\usepackage[utf8]{inputenc}
\usepackage{graphicx}
\usepackage{amssymb}
\usepackage{mathtools}
\usepackage{amsmath}
\usepackage{color}
\usepackage{rotating}
\usepackage{pdflscape}
\usepackage{array}
\usepackage{multirow}
\usepackage{tabularx}
\usepackage{booktabs}
\usepackage{cite}
\usepackage{threeparttable}
\usepackage{caption}
\usepackage{subcaption} 
\title[U Ant Sub-mm Detached Shell]{The Nearby Evolved Stars Survey: I. JCMT/SCUBA-2 Sub-millimetre detection of the detached shell of U Antliae}
\author[T.E. Dharmawardena et al.]{Thavisha E. Dharmawardena$^{1,2}$\thanks{tdharmawardena@asiaa.sinica.edu.tw}, Francisca Kemper$^{3, 1}$, Sundar Srinivasan$^{1,4}$,   
\newauthor
Peter Scicluna$^{1}$, Jonathan P. Marshall$^{1}$, Jan G. A. Wouterloot$^{5}$, Jane Greaves$^{6}$,   
\newauthor
Steven R. Goldman$^{7}$, Jacco Th. van Loon$^{8}$, Mikako Matsuura$^{6}$, Iain McDonald$^{9}$,
\newauthor
Jinhua He$^{10, 11, 12}$, Albert A. Zijlstra $^{9,13}$, Jes\'us A. Toal\'a$^{4}$, Sofia H. J. Wallstr\"om$^{1,14}$,
\newauthor
Hyosun Kim$^{15}$,  Alfonso Trejo$^{1}$, Paolo Ventura$^{16}$, Eric Lagadec$^{17}$, Martha L. Boyer$^{7}$, 
\newauthor
Tie Liu$^{15, 5}$, Gioia Rau$^{18, 19}$, Hideyuki Izumiura$^{20}$, Jan Cami$^{21, 22}$, Wayne Holland$^{23}$,
\newauthor
Olivia Jones$^{23}$, Hiroko Shinnaga$^{24}$,\\
$^{1}$ Academia Sinica Institute of Astronomy and Astrophysics, 11F of AS/NTU Astronomy-Mathematics Building, \\No.1, Sec. 4, Roosevelt Rd, Taipei 10617, Taiwan.\\
$^{2}$ Graduate Institute of Astronomy, National Central University, 300 Zhongda Road, Zhongli 32001, Taoyuan, Taiwan.\\
$^{3}$ European Southern Observatory, Karl-Schwarzschild-Str. 2, 85748 Garching b. M\"unchen, Germany \\
$^{4}$ Instituto de Radioastronom\'ia y Astrof\'isica, Universidad Nacional Aut\'onoma de M\'exico, Antigua Carretera a P\'atzcuaro \# 8701, \\ Ex-Hda. San Jos\'e de la Huerta, Morelia, Michoac\'an, M\'exico C.P. 58089. \\
$^{5}$ East Asian Observatory, 660 N A'ohoku Place, Hilo, Hawaii 96720, USA.\\
$^{6}$ School of Physics and Astronomy, Cardiff University, 4 The Parade, Cardiff CF24 3AA, UK. \\
$^{7}$ STScI, 3700 San Martin Drive, Baltimore, MD 21218 USA \\
$^{8}$ Lennard-Jones Laboratories, Keele University, ST5 5BG, UK. \\
$^{9}$ Jodrell Bank Centre for Astrophysics, Alan Turing Building, University of Manchester, Manchester, M13 9PL, UK. \\
$^{10}$ Key Laboratory for the Structure and Evolution of Celestial Objects, Yunnan Observatories, Chinese Academy of Sciences, \\
396 Yangfangwang, Guandu District, Kunming, 650216, P. R. China\\
$^{11}$ Chinese Academy of Sciences South America Center for Astronomy, China-Chile Joint Center for Astronomy, \\
Camino El Observatorio \#1515, Las Condes, Santiago, Chile \\
$^{12}$ Departamento de Astronom\'{i}a, Universidad de Chile, Casilla 36-D, Santiago, Chile\\
$^{13}$ Laboratory for Space Research, University of Hong Kong, Pokfulam Road, Hong Kong \\
$^{14}$ Institute of Astronomy, KU Leuven, Celestijnenlaan 200D bus 2401, 3001 Leuven, Belgium \\
$^{15}$ Korea Astronomy and Space Science Institute, 776, Daedukdae-ro, Yuseong-gu, Daejeon, 34055, Republic of Korea\\
$^{16}$ INAF, Osservatorio Astronomico di Roma, Via Frascati 33, Monte Porzio Catone, 00077 Roma, Italy \\
$^{17}$ Universit\'e C\^ote d’Azur, Observatoire de la C\^ote d’Azur, CNRS, Lagrange, France\\
$^{18}$ NASA/GSFC, Code 667, Goddard Space Flight Center, Greenbelt, MD 20071, USA \\
$^{19}$ CRESST II and Department of Physics, Catholic University of America, Washington, DC 20064, USA \\
$^{20}$ Okayama Branch Office, Subaru Telescope, NAOJ, NINS 3037-5 Honjo, Kamogata, Asakuchi, Okayama, 719-0232, JAPAN \\
$^{21}$ Department of Physics and Astronomy and Centre for Planetary Science and Exploration (CPSX), \\The University of Western Ontario, London, ON N6A 3K7, Canada\\
$^{22}$ SETI Institute, 189 Bernardo Ave, Suite 100, Mountain View, CA 94043, USA\\
$^{23}$ UK Astronomy Technology Centre, Royal Observatory, Blackford Hill, Edinburgh EH9 3HJ, UK\\
$^{24}$ Department of Physics and Astronomy, Graduate School of Science and Engineering, \\Kagoshima University
1-21-35 Korimoto, Kagoshima 890-0065 JAPAN\\
}

\begin{document}

\maketitle

\clearpage

\begin{abstract}

We present the highest resolution single-dish submillimetre observations of the detached shell source U Antliae to date. The observations were obtained at $450~\micron$ and $850~\micron$ with SCUBA-2 instrument on the James Clerk Maxwell Telescope as part of the Nearby Evolved Stars Survey. The emission at $850~\micron$ peaks at $40\arcsec$ with hints of a second peak seen at $\sim 20\arcsec$. The emission can be traced out to a radius of $56\arcsec$ at a $3\sigma$ level. The outer peak observed at $850~\micron$ aligns well with the peak observed at Herschel/PACS wavelengths. With the help of spectral energy distribution fitting and radiative transfer calculations of multiple-shell models for the circumstellar envelope, we explore the various shell structures and the variation of grain sizes along the in the circumstellar envelope. We determine a total shell dust mass of $(2.0 \pm 0.3) \times 10^{-5}$ M$_{\odot}$ and established that the thermal pulse which gave rise to the detached shell occurred 3500 $\pm$ 500 years ago.

\end{abstract}


\section{Introduction}
\label{Sec:Introduction}

In the final stages of stellar evolution stars expel their outer layers enriched with the products of nucleosynthesis into the interstellar medium (ISM). For intermediate-mass stars ($1M_{\odot} \leq M \leq 8M_{\odot}$), the majority of this mass loss occurs while on the Asymptotic Giant Branch (AGB) in a pulsation-enhanced, radiation-pressure driven wind \citep{Hofner2018}.

AGB stars are often treated as quasi-stable systems, without incorporating treatment of their evolution, while their winds are treated as spatially and temporally homogeneous outflows. The existence of complex structures such as elongations, detached shells and bipolar outflows \citep{Zijlstra2001, Olofsson2010, Ramstedt2011, Cox2012, Maercker2012}, indicates that the true mass-loss mechanisms are far more complex than commonly inferred. Particularly uncertain is the extent to which the stellar wind is enhanced in mass and/or momentum when the star undergoes a thermal pulse (He-shell flash). Thus, further observational constraints are required before we can statistically model mass loss from AGB stars accurately. By studying the extended dust emission and comparing it to constant-outflow models and detailed numerical simulations \citep[e.g][]{Bowen1991, Hofner2008}, we can study the properties of this non-uniformity and accurately determine time-variant mass-loss and dust-production rates and establish the properties of the grains that enter the ISM.

Of the variety of structures shown by AGB envelopes, detached dust shells are among the most striking features. They are thought to result from a period of strong mass loss due to a thermal pulse, during which the star may expand and brighten dramatically for a few centuries \citep{Willems_deJong1988-DetachedShell, Vassiliadis1994, Marigo2017}. According to predictions from evolutionary models \citep[e.g.,][]{Mattsson2007_DetShellHydro} the mass-loss rate during the thermal pulse is more than an order of magnitude greater than before the thermal pulse, mainly driven by the temporal increase in luminosity. A faster wind speed during this period means that the older, slower wind in front of the density-enhanced wind piles up at the shock interface into a shell \citep{Mattsson2007_DetShellHydro}. Such a detached shell source would be observed as a nearly symmetrical ring of gas and dust surrounding a near empty region around the star \citep{Zijlstra2001, Schoier2005}. 

Similar morphologies have been observed in a number of objects  \citep[e.g.][]{Olofsson1988, Olofsson1990, Izumiura1996, Izumiura1997, Izumiura2011}. After the thermal pulse the star's luminosity rapidly diminishes to below pre-pulse levels, and then gradually recovers. This would have led to a drop in mass-loss rate and wind speed soon after the increase in both of these quantities \citep{Steffen2000_DetShellHydro}. Because the effect of luminosity on mass-loss rate is greater than that on wind speed \citep[e.g.,][]{Eriksson2014, Goldman2017}, this reversal exacerbates the contrast in wind density in the wake 
of the shell.

U Antliae is a C-rich AGB star located at a distance of $268 \pm 39$ pc \citep{vanLeeuwen2007}. It is surrounded by a well-defined detached shell, estimated to have been expelled by the star $\sim 2800$ years ago \citep{Kerschbaum2010}. Independent scattered-light, $^{12}$CO low-J rotational line emission, mid-IR and far-IR observations of U Ant all show radically different structure, making this source rather unique. We summarise the published results in Table~\ref{Table:PublishedRadii} in the appendix and a schematic diagram showing their mean shell radii and FWHMs in Fig.~\ref{fig:Schematic}.

Optical scattered-light observations by \citet{Gonzalez2001, Gonzalez2003} reveal four sub-shells at $\sim$ 25$\arcsec$, 37$\arcsec$, 43$\arcsec$ and 46$\arcsec$ from the star (hereafter ss1, ss2, ss3, ss4), with the innermost two shells only tentative detections. They derive shell widths of $\sim 3\arcsec$, $\sim 6\arcsec$, $\sim 3\arcsec$ and $\sim 10\arcsec$ for ss1, ss2, ss3 and ss4 respectively. These authors find ss3 to be dominated by line-scattered light (i.e. resonance scattered light) instead of dust-scattered light indicating that ss3 is dominated by gas instead of dust.  
Follow-up observations by \citet{Maercker2010} also observed ss3 and ss4 in optical scatted light at $\sim 43\arcsec$ and $\sim 50\arcsec$ with shell widths of $\sim 2\arcsec$ and $\sim 7\arcsec$. They find ss3 to be fainter in dust-scattered light and brighter in line-scattered light. 
While features appear at the positions of ss1 and ss2 in their azimuthally-averaged surface-brightness profiles, 
the authors argue that they are a result of substructure in ss3 and ss4 projected towards the inner regions of the detached shell.

Observations of thermal dust emission, however, tell a different story. 
Mid-IR and far-IR observations from AKARI show that the surface brightness peaks at $\sim 41\arcsec$ \citep{Arimatsu2011}. However, they assume the double shell model (including peak radii and FWHMs) presented in \citet{Maercker2010} when analysing their data. \citet{Izumiura1997} et al., suggest the presence of two shells at $\sim 46\arcsec$ and $\sim 3'$ based on IRAS images. The latter has not been recovered by any other observations to date. 
Curiously, far-IR observations from Herschel/PACS appear to be dominated by ss3, peaking at $40\arcsec$ \citep{Kerschbaum2010, Cox2012}.  As described, in all other instances ss3 is observed to have very faint emission in dust continuum but very bright in gas emission. 

The shell has also been extensively observed in sub-mm CO lines. \citet{Olofsson1996} first mapped the gas shell in CO (1-0), (2-1) and (3-2), albeit at low resolution, locating the shell at $41\arcsec$ (width = $13\arcsec$).
The APEX $^{12}$CO(3-2) radial profile \citep{Maercker2010} clearly peaks at the location of ss3 and has a measured shell width of $\sim 2.6\arcsec$. High spatial resolution ($1.5\arcsec$) ALMA $^{12}$CO(2-1) and (1-0) observations by \citet{Kerschbaum2017} also only detect a single CO gas shell coinciding well with ss3. The CO shell is located at $42.5\arcsec$ from the central source and has a measured width of $\sim 5\arcsec$. The ALMA observations also show filamentary sub-structure within the gas shell.


As mentioned above, \citet{Maercker2010} suggest that ss3 and ss4 are real, while ss1 and ss2 are filamentary substructures of ss3 and ss4 projected against the inner regions of the detached shell. They show that the small distance between ss3 and ss4 and the corresponding time scales ($\sim 110$ yrs) suggest that these sub-shells could not have occurred due to multiple thermal pulses. The most likely scenario is a single thermal pulse $\sim 2800$ yrs ago gave rise to the detached shell following which a secondary mechanism shaped the single detached shell into the multiple sub-shells observed. 

A model for multiple shell formation in AGB and post-AGB stars was proposed by \citet{Simis2001a}. They suggest alternating dust and gas shells $200 - 400$ years apart formed as a result of dust and gas decoupling. In a similar vein for U Ant, \citet{Maercker2010} proposed the splitting of a single detached shell (located at the position of ss3) into two is an effect of gas-grain decoupling due to varying expansion velocities, resulting in a single, gas-rich sub-shell and a dust component at larger radii due to the higher expansion velocity. In this scenario the gas velocity slows down in the wind collision region while the dust sails through.

Should ss1 and ss2 be real, an as-of-yet unknown mechanism is required to explain their formation. One possibility is that instabilities at the interaction between the fast and slow winds may have created multiple shock fronts with dust decoupling in the swept back shock, resulting in ss1 and ss2 \citep{Gonzalez2001, Schoier2005, Kerschbaum2017}. The presence of filamentary structure in the gas-rich shell in the ALMA observation by \citet{Kerschbaum2017} provides evidence for a reverse shock. Another is that these shells could be a result of density and velocity modulations which took place during the thermal pulse \citep{Villaver2002, Maercker2010}.


While U Ant is well studied from the optical to the far-IR, only a few sub-mm continuum observations of the source exists. Archival observations obtained using the JCMT/SCUBA instrument (the predecessor to SCUBA-2) in 1997 (PI: Greaves) were never published until now, and do not clearly show the detached shell due to a low signal-to-noise ratio (see Appendix~\ref{sec:App:SCUBAobs}). U Ant was also part of a sample of three detached shell sources studied by \citet{Maercker_18_Properties} at $870~\micron$ using APEX/LABOCA. The authors report a sub-mm excess in the detached shell when comparing the observed fluxes to the output from radiative transfer models derived by combining data from the optical to the far-IR and extrapolating to the sub-mm. They measure an excess which is $2.3\pm0.3$ times greater than the model predictions. 

In this paper we present the highest angular-resolution ($13\arcsec$) submillimetre dust continuum detection of the detached shell of U Ant to date. The observation was obtained with the James Clerk Maxwell Telescope's (JCMT) Sub-millimetre Common-User Bolometer Array 2 \citep[SCUBA-2;][]{Holland2013} instrument, as part of the Nearby Evolved Stars Survey (NESS; Scicluna et al., \emph{in prep.}). 

Using this new sub-mm data, combined with archival Herschel/PACS data we study the dust properties and masses in this unique detached shell source. The analysis is carried out with the aim of reconciling the differences seen in the various types of observations. Using radiative transfer modelling we will evaluate whether our observations are consistent with the dust distribution over the multiple sub-shells as reported in the past. 

As part of the NESS data release, the raw SCUBA-2 data used in this paper will be available in the near future. The scripts and reduced data required to reproduce the analysis, figures and tables presented in this paper is available in figshare from \url{https://figshare.com/projects/UAnt_Submm/67421} under the project title \textit{UAnt\_Submm}.

\section{Observations and Data Reduction}
\label{Sec:Obs+DataReduction}

U Ant was observed on 18th of January 2018 as part of NESS (program ID: M17BL002) with SCUBA-2 on the JCMT. The observations were carried out at $450~\micron$ (beam FWHM = 7.9\arcsec) and $850~\micron$ (beam FWHM = 13\arcsec) using the CV-daisy scan pattern. The total observing time was 2.1 hours broken into four repeats. 

The data were reduced using the modified SCUBA-2 pipeline presented in \citet{Dharmawardena2018} via Starlink \citep{Currie2014} version 2018A. In general SCUBA-2 pipelines are built to handle bright point sources such as quasars or large extended structure such as molecular clouds. Compared to these evolved star circumstellar emission in the sub-mm is only marginally extended, leading to standard pipelines being unable to recover the circumstellar emission efficiently. Therefore in \citet{Dharmawardena2018} we developed a modified pipeline which can recover the marginally extended faint circumstellar emission optimally while suppressing artefacts.

\subsection{Removing CO(3-2) Contamination}
\label{sec:COcontam}

The wide bandwidth of the SCUBA-2 instrument ($790~\micron - 940~\micron$) contains the frequency of the CO(3-2) rotational transition. If this transition is strong enough, it may contaminate measurements of the continuum flux\footnote{\url{https://www.eaobservatory.org/jcmt/instrumentation/continuum/scuba-2/contamination/}} \citep{Drabek2012_SCUBA2_COcontam}. Therefore we carry out $^{12}$CO(3-2) subtraction on our SCUBA-2 $850~\micron$ observation. A full description of the methodology used to carry out this subtraction is presented in appendix~\ref{Sec:app_HARPred}.

We find a $\sim 30\%$ reduction in $850~\micron$ flux when the subtraction is carried out. This is on the upper end of the range reported by \citet{Drabek2012_SCUBA2_COcontam} when extreme cases are excluded. In the analysis to follow we use the $^{12}$CO(3-2) subtracted SCUBA-2 $850~\micron$ observation.

\subsection{Archival Herschel observations}
We combine the SCUBA-2 observations with Herschel/PACS $70~\micron$\ and $160~\micron$ imaging observations of U Ant as part of our analysis (FWHMs of $5.46 \times 5.76$ and $10.65 \times 12.13$ respectively). These data are a part of the Mass-loss of Evolved StarS (MESS) program \citep{Groenewegen2011} and are publicly available for download via the Herschel Science Archive\footnote{\url{http://www.cosmos.esa.int/web/herschel/science-archive}}. Here we use the Level 2.5 reduced products, the highest available pipeline-reduced data products calibrated using PACS calibration version \emph{PACS$\_$CAL$\_$77$\_$0}.

In addition to Herschel/PACS data we utilised the Herschel/SPIRE data from the MESS survey. The SPIRE beam FWHMs are $18\arcsec$ at $250~\micron$, $24\arcsec$ at $350~\micron$, and $42\arcsec$ at $500~\micron$, therefore the resolution of SCUBA-2 even at $850~\micron$ is at least a factor of 1.5 better. As the primary goal of this project is to analyse the sub-mm emission from the detached shell of U Ant, in order to preserve the SCUBA-2 resolutions we opt to not include the SPIRE data when carrying out spatial observational analysis. Therefore, we only use the SPIRE data for the SED analysis when carrying out radiative transfer modelling in Sec.~\ref{Sec:DustModelling}.

\section{Analysis and Results}
\label{Sec:Analysis+Results}

\subsection{Surface-brightness Profiles}
\label{sec:Methods:RadProfs}

\begin{figure*}
\centering
\begin{subfigure}{0.48\textwidth}
  \centering
  \includegraphics[width=1.1\textwidth]{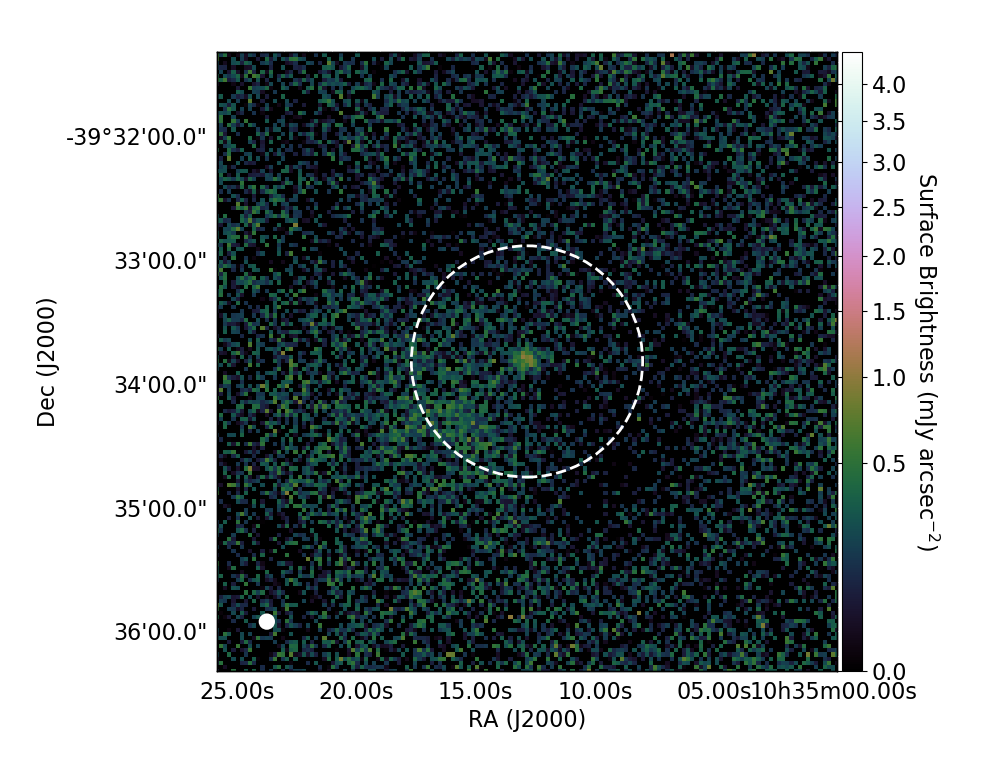}
  \subcaption{$450~\micron$}
  \label{fig:UAnt_450_Obs}
  \end{subfigure}
\begin{subfigure}{0.48\textwidth}
  \centering
  \includegraphics[width=1.1\textwidth]{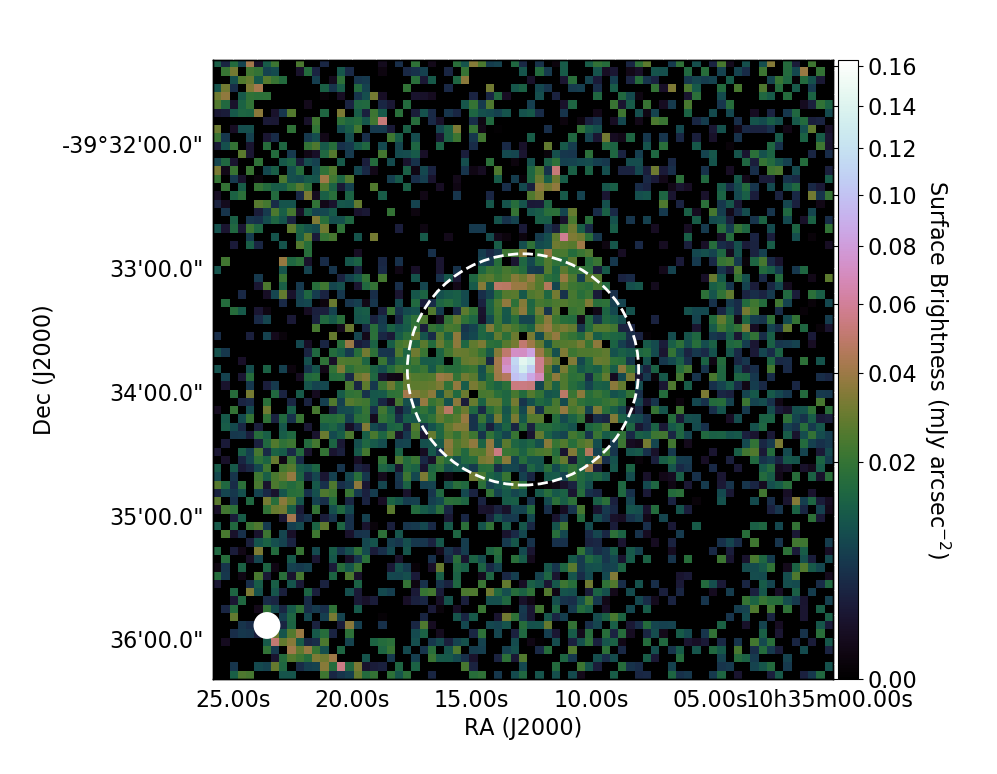}
  \subcaption{$850~\micron$}
  \label{fig:UAnt_850_Obs}
  \end{subfigure}
    \caption{(a) $450~\micron$ observation of U Ant (1 pix = $2\arcsec$); (b) CO (3-2) subtracted $850~\micron$ observation of U Ant (1 pix = $4\arcsec$). Dashed white circle: $3\sigma$ surface-brightness extent ($56\arcsec$) at $850~\micron$. Filled white circle in the bottom left corner: SCUBA-2 beam ($450~\micron$ beam FWHM = $7.9\arcsec$ and $850~\micron$ beam FWHM = $13\arcsec$)}
  \label{fig:UAnt_SCUBA-2_Obs}
\end{figure*}

The SCUBA-2 $450~\micron$ map has an RMS of 0.24 mJy arcsec$^{-2}$ and a pixel size of $2\arcsec$ (see Fig.~\ref{fig:UAnt_450_Obs}). Presented in Fig.~\ref{fig:UAnt_850_Obs}, the final CO subtracted $850~\micron$ map has an RMS of  0.02 mJy arcsec$^{-2}$ and a pixel size of $4\arcsec$. The $850~\micron$ image clearly shows a circumstellar envelope (CSE) extending over the full region of the detached shell of U Ant. 


\begin{figure*}
   \centering
  \includegraphics[width=0.8\textwidth]{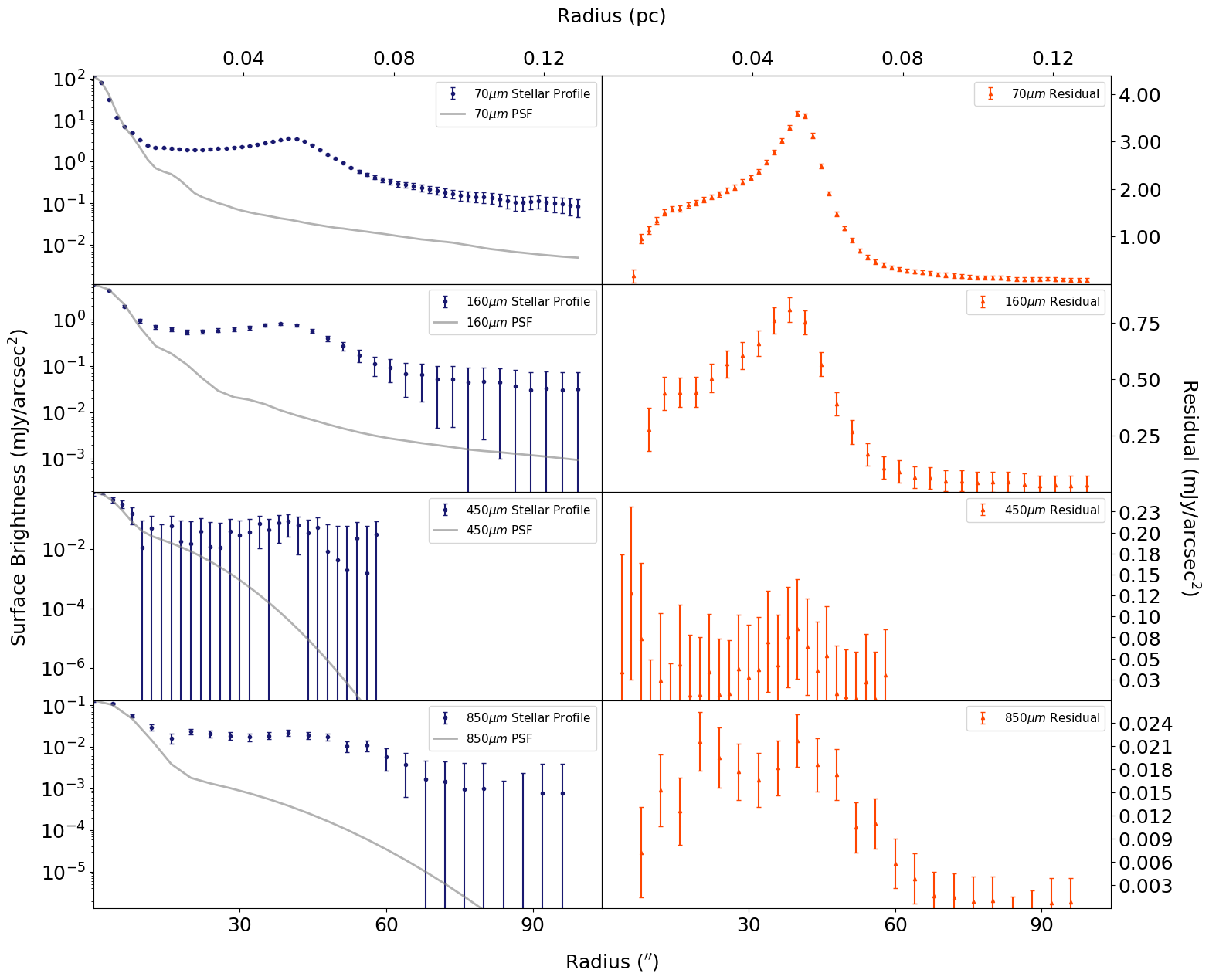}
   \caption{Surface-brightness and residual profiles of U Ant. Left hand panels: The blue dashed lines represent the source surface-brightness profiles and the grey solid lines represent the PSF profile of the instrument at the given wavelength; Right hand panels: The orange lines represents the PSF subtracted residual profiles for each wavelength.}
   \label{fig:UAnt_RadProf}
\end{figure*}

We derive the surface-brightness and PSF-subtracted residual profiles using the methods described in \citet{Dharmawardena2018} (see Fig.~\ref{fig:UAnt_RadProf}). 

The uncertainties on the residual profiles are the quadrature sum of the uncertainty on the PSF and the uncertainty on the radial profile:

\begin{equation}
    \sigma_{res}^{2} = \sigma_{SB}^{2} + \left[ \left(\frac{\sigma_{PSF}}{F_{PSF}}(r=0)\right) F_{PSF}\right]^{2},
\end{equation}

where $\sigma_{res}$ is the uncertainty on the residual profile, $\sigma_{SB}$ is the uncertainty on the surface brightness profile and $\sigma_{PSF}$ is the uncertainty on the PSF after it is scaled to the central peak pixel. The fractional uncertainty on the PSF (i.e.  $\frac{\sigma_{PSF}}{F_{PSF}}(r=0)$) is equal to the fractional uncertainty on the peak of the radial profile to which the PSF is scaled. The shape of the SCUBA-2 PSF is well known, therefore no significant uncertainty arises due its shape. As we align the PSF with 0.1 pixel precision, any effects due to misalignment are negligible.

The $850~\micron$ residual profile show a broad peak centred at $40\arcsec$. We also observe hints of an additional inner peak centred at $\sim 20\arcsec$. The outer maximum corresponds well to their Herschel/PACS counterparts. The  $850~\micron$ residual profile has a surface-brightness extent of $56\arcsec$ ($0.07 \pm 0.01$ pc) at $3\sigma$ detection limit (R$_{3\sigma}$), which is comparable to the R$_{3\sigma}$ we measure at both Herschel/PACS wavelengths. 

The $450~\micron$ profiles shows hints of emission from the detached shell once again at $\sim 40\arcsec$. However, the low signal-to-noise of the observation limits our ability to constrain this emission any further. 

Background-subtracted total fluxes (central source + extended component: $F_{\rm total}$) at $450~\micron$ and $850~\micron$ were measured to be $435 \pm 70$ mJy and $199 \pm 34$ mJy respectively. Fluxes at both wavelengths are derived using an aperture of $56\arcsec$ (R$_{3\sigma}$ at $850~\micron$) and a sky annulus from $80\arcsec$ to $120\arcsec$. The PSF-subtracted (or extended) component of the CSE accounts for $80\%$ of the total flux at $850~\micron$. 


\subsection{Shell Modelling
\label{Sec:All_Modelling}}

We carry out two sets of modelling in order to discern the detached-shell properties of U Ant in a step-by-step manner. The first of these interprets the extended emission observed in the far-IR and sub-mm in isolation (further described in Sec.\ref{Sec:SEDfitting}). Here we fit the four-point Spectral Energy Distribution (SED) at each radial point of the extended circumstellar envelope derived by combining the residual profiles (following the subtraction of the central point sources) at each of the four wavelengths. By fitting the thermal dust emission of the extended CSE at each radial point we derive the dust Temperature ($T$), Spectral index of dust emissivity ($\beta$ and dust mass column density ($\Sigma$).

Second, we carry out self-consistent radiative-transfer modelling of the entire system, i.e. star $+$ shell (further described in Sec.~\ref{Sec:DustModelling}). We do this in an attempt to compare structures suggested in the literature to the global SED and far-IR and sub-mm extended emission and potentially exclude some scenarios due to incompatibility with our observations of U Ant.

\vspace{1cm}

\subsubsection{Radial point-to-point Spectral Energy Distribution Fitting}
\label{Sec:SEDfitting}

By combining all four residual profiles we derive the Spectral Energy Distributions (SEDs) at each radial point.  These SEDs are fitted with a single-temperature blackbody model modified by an effective emissivity law \citep[e.g.,][]{Hildebrand1983, Gordon2014} using the method presented in Sec.~3.2 of \citet{Dharmawardena2018}. This results in radial profiles for the dust $T$, $\beta$ and $\Sigma$. 

As with \citet{Dharmawardena2018}, the input dust model consists of a mixture of $90\%$ amorphous carbon (optical constants from \citealt{Zubkoetal1996}) and $10\%$ silicon carbide (optical constants from \citealt{Pegourie1988}). The grain size distribution is as prescribed by \citet{Kim1994_DustProperties} (a power-law with an exponential falloff); where we use a minimum grain size of 0.01 $\mu$m and an exponential scale factor of 1 $\mu$m, with a power-law slope of $-3.5$. This results in an effective emissivity at $160\micron$ (${{\kappa_{\rm eff,160}^{S}}}$) of $26$ cm$^{2}$g$^{-1}$. We use this same dust model in the analysis throughout the entire paper in order to ensure consistency between the different types of modelling carried out.

The fitting is performed with the python package \textsc{emcee}, \citep{Foreman-Mackey2013}, which uses affine-invariant Markov Chain Monte Carlo (MCMC) algorithms to carry out Bayesian inference on the SEDs to the specified model, and find the most probable value for each parameter at every radial point. The $1\sigma$ uncertainties of the profiles are the central $68\%$ of the samples of the posterior generated by \texttt{emcee} with its median being used as the estimator.

We made several modifications to the SED fitting MCMC code presented by \citet{Dharmawardena2018} to suit this analysis. In particular, the limits in the $T$ prior are set to 20 K $< T <$ 300 K, with the inner temperature set to 1800 K. The $\beta$ limits are set to be $\beta > 1$. We find these modifications allow for better converged results for U Ant. 

As described in Sec. 4.2 of \citet{Dharmawardena2018}, the curvature of the fitted modified blackbody depends on both $\beta$ and $T$. The temperature is constrained by the peak of the SED at each individual radial point, thus by the Wien end of the SED \citep{Shetty2009}. Hence the $T$ profile is constrained by the far-IR PACS detections. The $\beta$ profile is constrained by the longer wavelength ($\lambda \geq 300\micron$) SCUBA-2 detections as it describes the Rayleigh-Jeans tail of the SED \citep{Doty1994, Shetty2009, Sadavoy2013}. The $\Sigma$ profile is constrained by either the PACS or the SCUBA-2 detections.

There is a known anti-correlation between $T$ and $\beta$ in all three-parameter modified blackbody models and the best way to overcome this degeneracy is to employ hierarchical Bayesian inference \citep{Kelly2012_ModelDegeneracies}. However as we lack the required sample size to carry out hierarchical Bayesian inference we use an informative prior on $\beta$ (probability distribution function of $\beta$ observed in M31). While it can not completely remove the degeneracy, it helps to minimise its impact.

\begin{figure}
    \centering
    \includegraphics[width=0.5\textwidth]{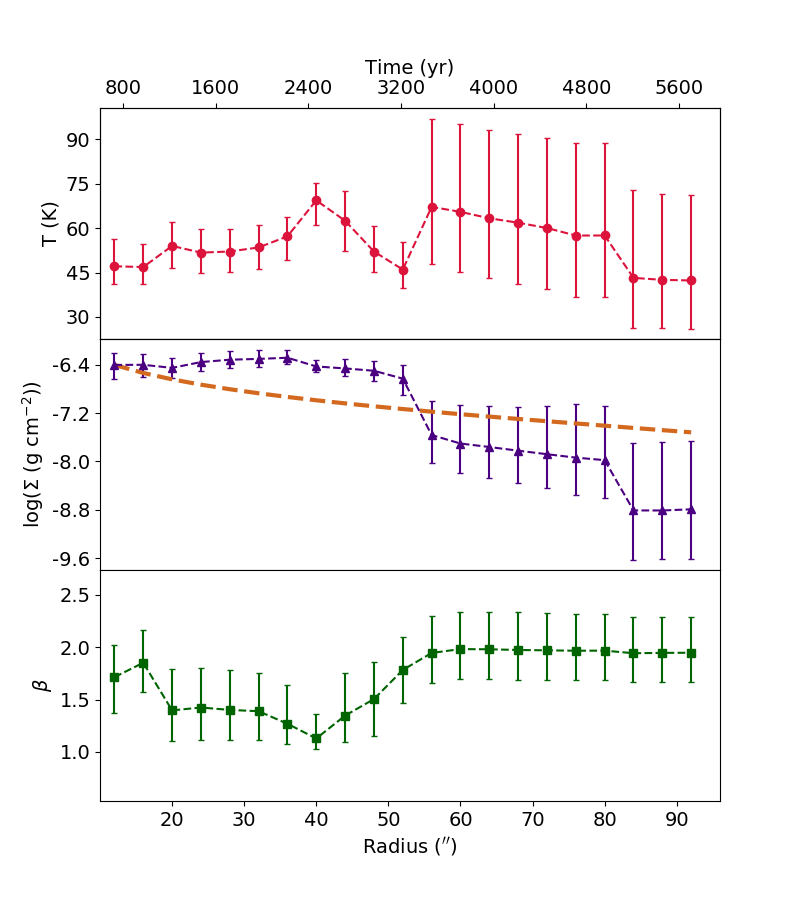}
    \caption{SED fitting results of U Ant; Top: Temperature ($T$) radial profile of U Ant; Middle:  Dust mass column density ($\Sigma$) radial profile. Orange dashed line represents the expected dust mass column density for a uniform mass-loss rate; Bottom: The spectral index of dust emissivity ($\beta$) profile of U Ant.}
    \label{fig:UAnt_T_B_D_profile}
\end{figure}

The resulting $T$, $\beta$ and $\Sigma$ profiles are presented in Figure \ref{fig:UAnt_T_B_D_profile}. Appendix~\ref{Sec:app_MCMC_Fitquality} shows an example of the median modified black body model at one radial point to illustrate the quality of the fit. 

By integrating over the $\Sigma$ profile from $12\arcsec$ to $56\arcsec$ we derive a total dust mass of $(2.0 \pm 0.3) \times 10^{-5}$ M$_{\odot}$ in the detached shell. This mass is assumed to be constant throughout the rest of this paper. This assumption may have an impact on the model SEDs and surface brightness profiles in Sec.~\ref{Sec:DustModelling}, but the very low optical depth of the shell means the effects will be negligible.

\subsubsection{Full Radiative Transfer Modelling}
\label{Sec:DustModelling}

In order to qualitatively determine the location of the far-IR/sub-mm dust emission we generate models using the python radiative-transfer package \textsc{Hyperion} \citep{Robitaille2011_Hyperion}. We compare the resulting model SEDs to the observed global SED of U Ant from optical to sub-mm (see Table~\ref{table:UAnt_SED_Table} in the appendix). Further, we qualitatively compare the resulting surface-brightness profiles to those derived from the SCUBA-2 and PACS observation. The best fit model SEDs and surface-brightness profiles allow us to narrow down the most likely CSE structure of U Ant. 


We choose six different model scenarios based on our observations and past literature reports. Since four distinct sub-shells have been reported we have experimented with scenarios that put all of the dust that we have measured in the radial SED fitting (Sec.~\ref{Sec:SEDfitting}) in one of each of these shells. For the fourth shell, two different radii were reported by \citet{Maercker2010} and \citet{Kerschbaum2010} using dust continuum observations. Therefore in total we arrive at five distinct model scenarios: Mss1; Mss2; Mss3, Mss4-M2010; Mss4-K2010.  A sixth scenario sees all dust distributed between the four sub-shells with the mass distribution determined from literature \citep[Mfourshells;][]{Gonzalez2001, Gonzalez2003, Maercker2010, Maercker_18_Properties}. The input parameters for the individual models are presented in Table~\ref{Table:Model_Parameters}. While the detached shell has an expansion velocity of 20.5 km s$^{-1}$ \citep{DeBeck2010} we do not use this as input parameter as we have placed the sub-shells in their correct positions our static models. 

In the case of U Ant, the angular resolution of the JCMT/SCUBA-2 at $850~\micron$ is comparable to the distance between the innermost and the outermost sub-shells. Therefore we are only able to test the extreme scenarios presented above. Further exploration of the distribution of dust and finer shell structure is not informative as we are unable to meaningfully constrain the free parameters.

\input{Tables/Model_Parameters.tex}
    
To derive the appropriate input synthetic stellar photosphere and its parameters required as input to Hyperion (e.g., stellar luminosity and effective temperature) we fit the observed global SED of U Ant from optical -- mid-IR ($0.5 - 10~\micron$) to the COMARCS stellar photosphere model grid \citep{Aringer2009}. Past reports have found that the optical depth of the detached shell of U Ant is very low from optical -- sub-mm \citep[e.g.,][]{Kerschbaum2010}  confirming that the optical -- mid-IR SED of U Ant is unaffected by the detached shell. Therefore the parameters derived are also unaffected by the detached shell, with the central stellar emission dominating in this wavelength range.

The input parameters to all models are as follows: 

\begin{itemize}
    \item Total shell mass: $(2.0 \pm 0.3) \times 10^{-5}$ M$_{\odot}$ (see Sec.~\ref{Sec:SEDfitting});
    \item Expansion velocity of present day mass loss: 4.5 km s$^{-1}$ \citep{Kerschbaum2017};
    \item An inverse-square dust density distribution in the detached shell/sub-shells.
    \item Stellar luminosity: 7000 L$_{\odot}$;
    \item Stellar effective temperature: 2600 K;
    \item stellar surface gravity: $\log{(g~[{\rm cm~s}^{-2}])} = -0.5$;
\end{itemize}

The derived synthetic stellar photosphere parameters are consistent with the study by \citet{DiCriscienzo2016}, who find that stars of metallicities typical of the solar neighbourhood, and mass in the range $1.2-2~M_{\odot}$, reach the C-star stage with luminosities and temperatures similar to the best-fit parameters given above (see Fig.6 in \citet{DiCriscienzo2016}).

We use the same dust composition as described in Sec.~\ref{Sec:SEDfitting}. We do not include an underlying contribution from mass-loss pre and post thermal pulse. The COMARCS model  provides a good fit to the optical -- mid-IR component of the observed global SED, implying that the present day mass-loss provides virtually no contribution to the thermal dust emission.

\begin{figure*}
    \centering
    \includegraphics[width=0.8\textwidth]{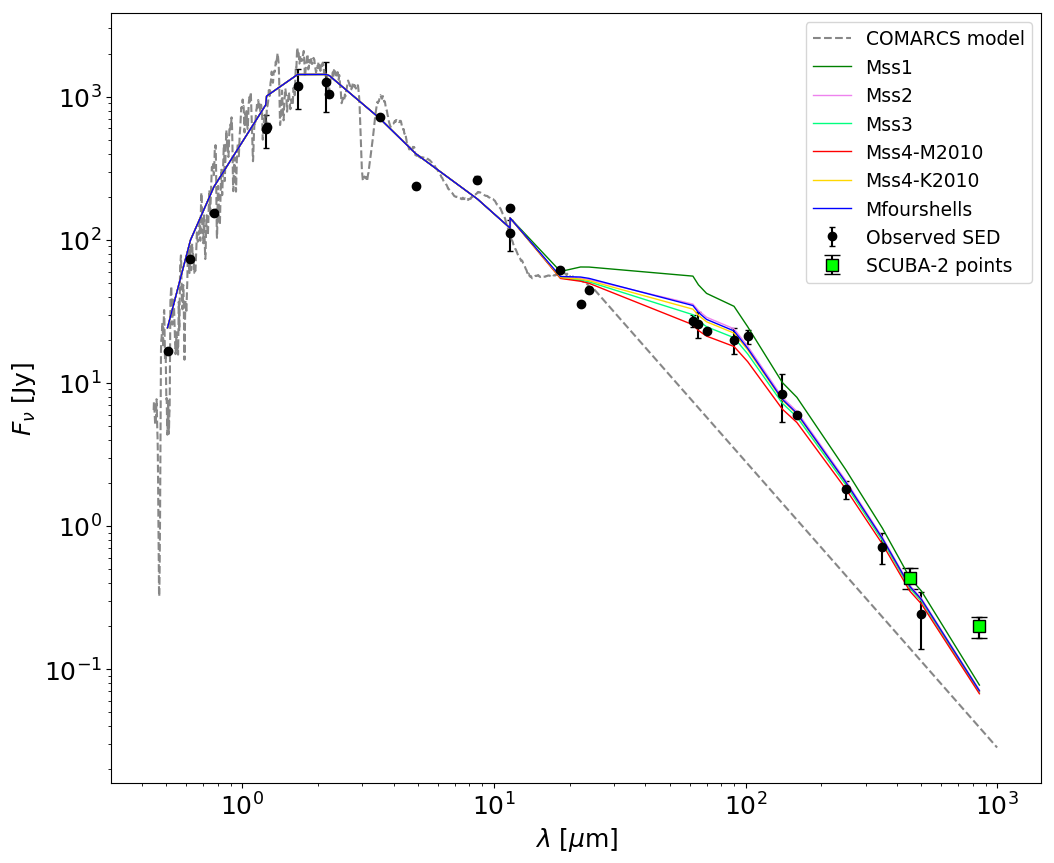}
    \caption{Comparison of model SEDs (synthetic photometry points connected by lines) with the observed SED (see Table~\ref{table:UAnt_SED_Table}). Black dots: observed SED; Grey dotted line: COMARCS model; Bright green dots: SCUBA-2 points; Dark green line: Mss1; Violet line: Mss2; Light green line: Mss3; Red line: Mss3-M2010; Gold line: Mss3-K2010; Blue line: Mfourshells.}
    \label{fig:ModelSEDs}
\end{figure*}

\begin{figure*}
    \centering
    \includegraphics[width=0.8\textwidth]{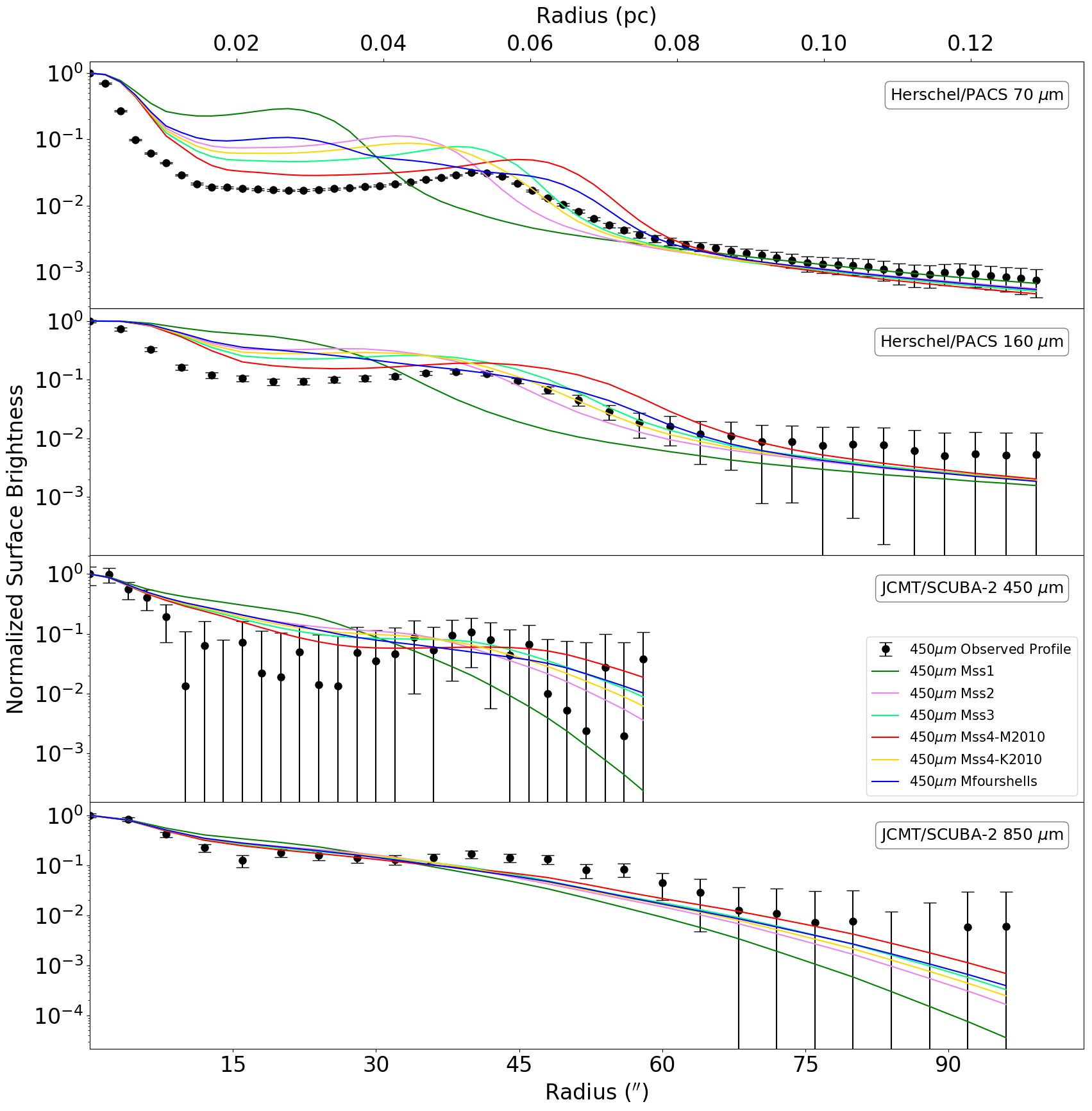}
    \caption{Comparison of model surface-brightness profiles with the observed. The legend shown in the SCUBA-2 $450~\micron$ panel applies to all other panels too. From top to bottom the plots show $70~\micron$, $160~\micron$, $450~\micron$ and $850~\micron$ respectively. Black dots: Observed surface-brightness profiles; Dark green line: Mss1; Violet line: Mss2; Light green line: Mss3; Red line: Mss3-M2010; Gold line: Mss3-K2010; Blue line: Mfourshells.}
    \label{fig:ModelRadProfs}
\end{figure*}

The resultant model SEDs are plotted along with the observed SED in Fig~\ref{fig:ModelSEDs}. The model surface-brightness profiles at each wavelength are shown along with the observed surface-brightness profiles in Fig~\ref{fig:ModelRadProfs}. The chi-squared values per observed data point ($\chi^{2}_{p}$) of the models when compared to both the observed SED and surface-brightness profiles are presented in Tab.~\ref{Table:RT_RedChiSq}.

\input{Tables/RT_RedChiSq.tex}

\section{Discussion} 
\label{Sec:Discussion}

\subsection{Surface-brightness emission}
\label{Sec:SurfaceBrightness}

We find that approximately $80\%$ of the total flux is emitted from the extended component in all four of the PACS and SCUBA-2 observations (see Sec.~\ref{sec:Methods:RadProfs}). This result is $\sim 25\%$ larger than the average reported in \citet{Dharmawardena2018}, as expected for a bright detached shell source. 

The peaks at $70~\micron$, $160\micron$ and the outer peak at $850~\micron$ align well at $40\arcsec$, despite the large difference in resolution, and match the weak emission at $450~\micron$. Additionally, \citet{Kerschbaum2010} report the same peak intensity radius, also using the MESS Herschel/PACS observations at $70~\micron$ and $160\micron$. Interestingly, none of the scattered-light peaks correspond to this sub-mm peak and it is located between ss2 and ss3. Further, the inner broad peak observed only at $850~\micron$ also does not correspond to any of the scattered light sub-shells and is located interior to ss1.

The aligned peaks in the SCUBA-2 and PACS profiles seen in Fig.~\ref{fig:UAnt_RadProf} point towards the presence of a single dust shell, with a far-IR/sub-mm peak at $\sim 40\arcsec$. We expect the peak is somewhat smeared by the beam sizes of the corresponding instruments. Such a single dust shell is consistent with the single gas-rich shell reported by \citet{Kerschbaum2017} using ALMA observations. \citet{Kerschbaum2017} suggests the gas-rich shell to have strong gas and dust coupling resulting in the peaks observed in both the CO observations and dust continuum observations. 

However we note another reason for the lack of distinct sub-shells could be the resolution of our observations. The separation between sub-shells is comparable to the FWHM of the SCUBA-2 beam ($13\arcsec$) resulting in the merging of shells in the observation. Nevertheless, if this were the case the multiple sub-shells should most likely have been visible in the Herschel/PACS $70\micron$ observations which has a much smaller beam size of $6\arcsec$, the approximate width of the shells in scattered light.

The additional emission peak at $\sim 20\arcsec$ seen at $850~\micron$ cannot be due to thermal dust emission, as the lack of emission at this radius at shorter wavelengths would require the dust to have a temperature of $\sim 4$ K (according to Wien's Law) which is unphysically low for dust so close to the star. Large or more amorphous grains radiate more efficiently at longer wavelengths \citep{Testi2014_GrainSize}, an effect that is expressed with low $\beta$ value, but no evidence for such an effect is found in the radially-derived $\beta$ values in Sec.~\ref{Sec:RadialDustProperties}. It thus remains unclear what the source of emission for this component is.


Following this, the inner brightness peak at $850~\micron$ is most likely a result of projection effects due to emission from filament-like/clumpy substructure within the single shell, resulting in discernible structure in the inner region of the CSE when projected against the plane of the sky. This is consistent with suggestions by \citet{Maercker2010} and the gaseous filamentary structure observed in past ALMA observations \citep{Kerschbaum2017}. The substructure likely possesses different grain properties to that of the overall average detached shell, hence causing it to appear only at $850~\micron$ and not at the PACS wavelengths. We discuss this further in Sec.~\ref{Sec:RadialDustProperties} by analysing the $T$, $\Sigma$ and $\beta$ profiles. 

$R_{3\sigma}$ at $850~\micron$ ($56\arcsec$) coincides well with the outer edge of ss4 detected by \citet{Gonzalez2001, Gonzalez2003} and \citet{Maercker2010} in scattered light. This is also consistent with the outer-most shell in mid-IR previously reported by \citet{Arimatsu2011}. This radius lies within the interquartile range measured for the sample of fifteen evolved stars by \citet{Dharmawardena2018}. 

Assuming a detached-shell expansion velocity of 20.5 km s$^{-1}$ \citep{DeBeck2010} and a distance of $268 \pm 39$ pc \citep{vanLeeuwen2007}, we trace the circumstellar shell out to a look-back age of $3500 \pm 500$ yr at $850~\micron$ at the $3\sigma$ level. This age is comparable to the ages obtained by \citet{Maercker2010} and \citet{Kerschbaum2017}.

\subsection{Radial variation in dust properties}
\label{Sec:RadialDustProperties}

In Figure~\ref{fig:UAnt_T_B_D_profile} we present the $T$, $\Sigma$ and $\beta$ SED fitting results calculated using \textsc{emcee}. The innermost $\sim 10\arcsec$ region of all three profiles is compromised by PSF-subtraction effects on the residual profile, and are therefore not included in the analysis or the figure. 

All three profiles are well constrained from $12\arcsec - 56\arcsec$, i.e. up to the $R_{3\sigma}$ radius at $850\micron$. The temperature profile peaks at $40\arcsec$, aligning well with the peaks observed in surface-brightness residual profiles. The overall weighted-averaged temperature is $54 \pm 2$ K within the region with constraints. The weighted-average dust temperature of the shell agrees well with the dust temperature reported in \citet{Schoier2005} using radiative transfer modelling. The temperature is also consistent with that expected for dust grain heating by the interstellar radiation field (ISRF). This suggests grains in this regions are heated by the same uniform ISRF and hence posses a similar temperature, giving rise to a single-temperature dust component. 

As shown by the middle panel in Fig.~\ref{fig:UAnt_T_B_D_profile}, the radial variation of $\Sigma$ clearly deviates from the uniform and constant mass-loss model overlaid in brown. The constant mass-loss model here is calculated by projecting a $r^{-2}$ density distribution from 3-dimensions to 2-dimensions. From $\sim 12\arcsec - 56\arcsec$ the $\Sigma$ profile follows an overall flat profile with no discernible peaks to indicate the presence of the detached shell. This could be due to interference of the substructure within the detached shell. Line-of-sight confusion -- a result of the substructure -- could cause the dust mass to appear spread evenly throughout the CSE when using this method. Therefore while we are able to estimate the outer radius of the detached shell using these parameter profiles, we are unable to identify the inner region. We see a sharp decrease in $\Sigma$ following this, indicating that approximately $3500 \pm 500$ ago there was an event of high mass injection to the CSE, i.e. the thermal pulse which gave rise to the detached shell.

The integrated dust mass from $12\arcsec - 56\arcsec$, $(2.0 \pm 0.3) \times 10^{-5}$ M$_{\odot}$ (statistical uncertainty only), is $\sim 3$ times smaller than that reported for ss4 by \citet{Maercker2010} based on optical scattered light. Given the likely uncertainties in measuring dust masses from scattered light these two measurements are probably consistent. The derived dust mass is consistent with those reported by \citet{Schoier2005, Arimatsu2011} and \citet{Maercker_18_Properties}.

By studying the upper and the middle panels in Fig.~\ref{fig:UAnt_T_B_D_profile}, we see that the $T$ and $\Sigma$ are prior dominated from $\sim 80\arcsec$ outwards. The region immediately ahead of the thermal pulse (i.e. $56\arcsec - 80\arcsec$) suggests there may be emission from pre-thermal pulse material below the 3$\sigma$ level likely observed by Herschel/PACS. 

The final panel in Fig.~\ref{fig:UAnt_T_B_D_profile} depicts the radial $\beta$ variation of the detached shell. The variation of $\beta$ up to $56\arcsec$ is a direct indicator that the grain properties vary radially. The difference from the canonical value of $\beta$ for ISM dust \citep[e.g.][]{PlanckCollaboration2014} demonstrates that there is no substantial contribution from a swept up ISM dust component. This supports arguments that the shell arises from a variation in mass loss rather than an interaction between the wind and the ISM \citep{Wareing2007}. Given the uncertainties it is difficult to pinpoint the location of the changes in $\beta$. The region beyond $56\arcsec$ is prior dominated as the best-fit value of $\beta$ is dictated by the SCUBA-2 data, which no longer provide strong constraints beyond this radius.

The broad peak at $16\arcsec$ ($\beta$ = 1.85) is aligned well with the inner residual profile peak at $850~\micron$, indicating that grains in this region are different to that of the rest of the shell, i.e. within the substructure of the shell. The dip at $40\arcsec$ ($\beta$ = 1.1) is consistent with the peaks in both the temperature and residual profiles. The minimum and maximum values of $\beta$ observed are intermediate to those expected for amorphous carbon and graphite \citep{Mennella1998, Draine2016}. 

\subsection{Self-consistent dust radiative transfer modelling}
\label{Sec:ModellingDiscussion}

The model SEDs in Fig.~\ref{fig:ModelSEDs} are indistinguishable from one another at wavelengths shorter than $11.6~\micron$, reproducing the general trend in observed global SED and the COMARCS model from the optical to mid-IR. From $11.6~\micron$ onwards the SEDs of all but the Mss1 align well with the observations up to $500~\micron$.

As expected given previous reports \citep{Kerschbaum2010, Maercker2010} and as seen in Fig.~\ref{fig:ModelSEDs}, the scenario of only an inner shell is unlikely. This is consistent with the largest $\chi^{2}_{p}$ value being derived for this model rendering it the least likely. This supports the suggestion by \citet{Maercker2010} that ss1 is an artefact. For a similar reason the model with all the dust in ss2 also does not provide a good fit. Given that the models of ss1 and ss2 do not fit well, it is expected for the Mfourshells to also not fit well. The reason for this is likely related to the temperatures of dust at these distances, which is too warm to reproduce the observed FIR emission.

Interestingly having all dust in ss3 best reproduces the SED with having all dust in the outermost shell (M2010) following a close second. Further, having all dust in the K2010 variety of the outermost shell has a much larger $\chi^{2}_{p}$ making it also unlikely (most likely due to warm dust in the inner region overlapping with ss2). Therefore it is possible that the detached shell of U Ant extends from $41\arcsec - 54\arcsec$. Previous reports of shell 3 having little-to-no dust may be premature, agreeing with the peak of the Herschel/PACS radial profiles, which peak at the location of ss3 where the gas emission peaks.

We note that similar to \citet{Maercker_18_Properties}, none of the models reproduce the flux at $850~\micron$, however they did not account for the contribution of CO(3-2) to the $850~\micron$ flux. Our analysis -- including CO subtraction -- still produces an excess in flux $\sim 3$ times the model predictions. 
This could be due to the difference in grain properties in the substructure only visible at $850~\micron$. Further exploration into dust properties (e.g.: size, shape and/or composition) and emission mechanisms (e.g.: spinning dust grains) may help understand this effect. Longer wavelength observation at e.g. 1.1 and 1.3 mm from ALMA or the LMT will contribute towards confirming the presence and shape of this excess. 

As seen in Fig.~\ref{fig:ModelRadProfs}, at $70~\micron$ and $160~\micron$ none of the model surface-brightness profiles reproduce the observed surface-brightness profile up to $\sim 40\arcsec$. While not aligning well, having all dust in ss3 best reproduces the shape of the observed profile (only scaled up) once again agreeing with Herschel/PACS observations peaking within ss3. It is followed closely by Mss4-M2010 which is the second-best-fitting model, providing further evidence of the dust emission being within $41\arcsec - 54\arcsec$ (with complex structure and varying dust components as suggested below within the shell).

In contrast to the PACS data, the inner regions of the observed SCUBA-2 profiles are well reproduced by all models and begin to deviate only at $\sim 30\arcsec$. However, at $450~\micron$ even with the lowest $\chi^{2}_{p}$ results, the low significance of the flux prevents meaningful conclusions based on the current $450~\micron$ data. 

Results derived from the $850~\micron$ profiles are significantly different compared to the other three wavelengths, with only a small deviation between the $\chi^{2}_{p}$ values. Mss1, with the largest $\chi^{2}_{p}$ values at the other wavelengths (and can therefore easily be ruled out as a possibility), is the lowest $\chi^{2}_{p}$ at $850~\micron$. This could point towards the presence of a different dust component emitting at this wavelength at smaller projected separations from the star, suggesting the presence of shell-substructure projected inwards.


We suggest two reasons for our inability to reproduce the inner regions of the PACS profiles based on the fact that the central source is much more compact than the models predict. The first is that a present day MLR governed by a steeper density power law ($< -2$) needs to be applied. This would indicate that the present day MLR is increasing once more, consistent with the gradual recovery of the luminosity and MLR in the aftermath of the thermal pulse. The second is that there is a cut off in the density distribution as a result of the fast moving thermal pulse wind sweeping up the pre-thermal pulse mass loss essentially leaving a \textit{cavity} behind it. In order for the central component to appear point like at PACS $70~\micron$ (the smallest beam FWHM: $5.46\arcsec \times 5.76\arcsec$) it must be no more than $\sim 1/2$ beam FWHM \citep[e.g.: Table 2 in][]{Miettinen2015}. Higher resolution observations (e.g.: from ALMA) are required to probe this region. 

These scenarios become less significant at longer wavelengths as the beam size increases essentially smearing out the emission. We therefore do not have a reasonable explanation as to why the SCUBA-2 profiles begin to deviate following the peak of the detached shell.

\section{Conclusions}
\label{Sec:Conclusion}

We present the highest resolution sub-mm observations of the detached shell of U Ant at $850~\micron$ to date using JCMT/SCUBA-2. The detached shell is clearly detected at $850~\micron$ and marginally at $450~\micron$. It has a $3\sigma$ extent at $850~\micron$ of $56\arcsec$ ($0.07 \pm 0.01$ pc), consistent with past publications. The PSF-subtracted residual profile at $850\micron$ shows two peaks centred at $\sim 20\arcsec$ and at $40\arcsec$. The outer peak is aligned well with the peaks of the Herschel/PACS residual profiles at $70~\micron$ and $160~\micron$ and the weak emission at $450~\micron$. Therefore the well aligned peaks at all four wavelengths can be explained by the presence of a single shell. Hence, the inner residual peak observed at $850~\micron$ may be the result of substructure within the same shell, visible only at this longer sub-mm wavelength due to a difference in grain properties between the average shell and the sub-structure.

From radial point-to-point SED fitting we derive profiles for $T$, $\Sigma$ and $\beta$. The $T$ profile has a weighted averaged temperature of $54 \pm 2$ K (between $12\arcsec - 56\arcsec$) and is consistent with dust heated by ISRF. The sudden mass-loss increase in the $\Sigma$ profile at $56\arcsec$ points to the time of the thermal pulse which gave rise to the detached shell. We calculate it to have occurred approximately $3500 \pm 500$ yr in the past. By integrating the $\Sigma$ profile observed we estimate a total shell dust mass of $(2.0 \pm 0.3) \times 10^{-5}$ M$_{\odot}$. We see hints of pre-thermal pulse mass loss in the $\Sigma$ from $\sim 56\arcsec - 80\arcsec$. Radial variations in the dust properties would explain the variations seen in the $\beta$ profile; this would indicate the presence of dust grains with $\beta$ values intermediate to amorphous and graphitic carbon.


In all six of the model scenarios tested using radiative transfer modelling we are unable to reproduce the flux observed at $850~\micron$. This excess may be due to the substructure discussed above, however further analysis is required to better understand it. Resolved continuum observations at 1.1 and 1.3 mm will reveal the nature of this excess. 

We find that the two best-fitting models to both the global SED and the observed surface-brightness profiles are that of all the dust concentrated in sub-shell three and all the dust concentrated in sub-shell four with both having very similar $\chi^{2}_{p}$ values. This is in disagreement with existing literature which claims that sub-shell three has little-to-no dust. The detached shell of U Ant thus likely extends from $\sim 41\arcsec - 54\arcsec$.

At PACS wavelengths none of the models reproduce the inner $\sim 40\arcsec$ of the shell. At SCUBA-2 wavelengths the exact opposite occurs. The SCUBA-2 $850~\micron$ observation is best reproduced by the model scenario assuming a single inner shell which was previously ruled out. 

Two scenarios could give rise to the models being unable to reproduce the inner regions of the PACS wavelengths: (i) A present day MLR governed by a steep power law needs to be applied since the present day MLR and luminosity maybe increasing as a result of post-thermal pulse recovery; (ii) A cavity is formed as a result of the fast wind arising from the thermal pulse. Both these scenarios result in a highly compact central component which can not be constrained with current observations. These reasons become less significant at SCUBA-2 wavelengths as the beam size increases. Therefore we are unable to understand the reasoning for the deviations seen in SCUBA-2. 

Comparing the observations, derived surface-brightness profiles, dust parameter profiles and the radiative transfer modelling, we suggest that the detached shell of U Ant is a single dust shell. Filamentary/clumpy substructure similar to that reported by \citet{Kerschbaum2017}, within this shell appears closer to the central star due to line-of-sight projection effects. The grain properties of the substructure are different to that of the overall shell. 


Continuum observations from ALMA and SOFIA/FORCAST along with complex 3-D hydrodynamical modelling in the future could help resolve the variations observed in the model comparisons. The high resolution observations will constrain the dust radii and the inner dust components allowing us to observe any cut off and therefore the correct dust density distribution. Further, polarimetric imaging observation in the sub-mm will help narrow down the dust grain shape and size thus constraining the properties of the substructure within the detached shell.

\section*{Acknowledgements}
We thank the anonymous referee for their careful reading of the manuscript.
TED wishes to thank Prof. Chung-Ming Ko at NCU for his support of this project.
We are grateful to Matthias Maercker for the engaging discussions on the topic.
This research has been supported under grants MOST104-2628-M-001-004-MY3 and MOST107-2119-M-001-031-MY3 from the Ministry of Science and Technology of Taiwan, and grant  AS-IA-106-M03 from Academia Sinica.
IM acknowledges support from the UK Science and Technology Facilities Council under grant ST/P000649/1. 
JHH is supported by the NSF of China under Grant Nos. 11873086 and U1631237, partly by Yunnan province (2017HC018), and also partly by the Chinese Academy of Sciences (CAS) through a grant to the CAS South America Center for Astronomy (CASSACA) in Santiago, Chile. 
MM is supported by an STFC fellowship.  

The James Clerk Maxwell Telescope is operated by the East Asian Observatory on behalf of The National Astronomical Observatory of Japan; Academia Sinica Institute of Astronomy and Astrophysics; the Korea Astronomy and Space Science Institute; the Operation, Maintenance and Upgrading Fund for Astronomical Telescopes and Facility Instruments, budgeted from the Ministry of Finance (MOF) of China and administrated by the Chinese Academy of Sciences (CAS), as well as the National Key R\&D Program of China (No. 2017YFA0402700). Additional funding support is provided by the Science and Technology Facilities Council of the United Kingdom and participating universities in the United Kingdom and Canada.
\emph{Herschel} is an ESA space observatory with science instruments provided by European-led Principal Investigator consortia and with important participation from NASA.
In addition to software cited above, this research made use of the \textsc{Scipy} \citep{Scipy2001} and \textsc{Astropy} \citep{Astropy2018} python packages.
This research used the facilities of the Canadian Astronomy Data Centre operated by the National Research Council of Canada with the support of the Canadian Space Agency. This research also made use of the Canadian Advanced Network for Astronomical Research  \citep[CANFAR,][]{Gaudet2010-CANFAR}.
This work has made use of data from the European Space Agency (ESA) mission
{\it Gaia} (\url{https://www.cosmos.esa.int/gaia}), processed by the {\it Gaia}
Data Processing and Analysis Consortium (DPAC,
\url{https://www.cosmos.esa.int/web/gaia/dpac/consortium}). Funding for the DPAC
has been provided by national institutions, in particular the institutions
participating in the {\it Gaia} Multilateral Agreement.


\bibliographystyle{mnras}
\bibliography{UAnt_DetachedShell_Bib}

\input{Appendix.tex}

\end{document}

%% file: Tables/Model_Parameters.tex
\begin{table}
  \centering
  \caption{Input parameters for individual models}
    \begin{tabular}{llll}

\hline
\hline  

 \multirow{1}{*}{Model} &  Sub-shell   & \multicolumn{1}{l}{Shell Radius} & \multicolumn{1}{l}{$\%$ of total} \\
 &&&\multicolumn{1}{l}{dust mass}\\
 
\hline          

\vspace{0.1cm}
Mss1 & ss1 & $23.5\arcsec - 26.5\arcsec$ & $100\%$ \\

\vspace{0.1cm}
Mss2 & ss2 & $34\arcsec - 40\arcsec$ & $100\%$ \\

\vspace{0.1cm}
Mss3 & ss3 & $41.5\arcsec - 44.5\arcsec$ & $100\%$ \\

\vspace{0.1cm}
Mss4-M2010 & ss4 & $46.5\arcsec - 53.5\arcsec$ & $100\%$ \\ 

\vspace{0.1cm}
Mss4-K2010 &  ss4 & $34\arcsec - 46\arcsec$ & $100\%$ \\

Mfourshells &  ss1 & $23.5\arcsec - 26.5\arcsec$ & $22.5\%$  \\
& ss2 & $34\arcsec - 40\arcsec$ & $22.5\%$ \\
& ss3 & $41.5\arcsec - 44.5\arcsec$ & $5\%$ \\
& ss4 &  $46.5\arcsec - 53.5\arcsec$ & $50\%$ \\

\hline
     
    \end{tabular}%
  \label{Table:Model_Parameters}%

\end{table}%

%% file: Tables/RT_RedChiSq.tex
\begin{table*}
  \centering
  \caption{$\chi^{2}_{p}$ comparison between modelled and observed SED and surface brightness profiles.}
    \begin{tabular}{llllll}

\hline
\hline  

 \multirow{2}{*}{Model} & \multirow{1}{*}{SED} & \multicolumn{4}{l}{Surface Brightness Profiles $\chi^{2}_{p}$}\\
          &    $\chi^{2}_{p}$   & \multicolumn{1}{l}{$70~\micron$} & \multicolumn{1}{l}{$160~\micron$} & \multicolumn{1}{l}{$450~\micron$} & \multicolumn{1}{l}{$850~\micron$} \\
\hline          
          &       &       &       &       &  \\
          
Mss1 & 20596 & 6635 & 111 & 1 & 16 \\

Mss2 & 2259 & 610 & 21 & 1 & 18 \\

Mss3 & 635 & 148 & 15 & 1 & 19 \\

Mss4-M2010 & 687 & 258 & 19 & 1 & 19 \\ 

Mss4-K2010 &  1224 & 297 & 16 & 1 & 18 \\

Mfourshells &  1722 & 569 &  21 & 1 & 18 \\

\hline
     
    \end{tabular}%
  \label{Table:RT_RedChiSq}%

\end{table*}%


%% file: Appendix.tex
\appendix
\label{Sec:Appendix}

\section{Published Shell Radii and Schematic Diagram of U Ant}
\label{Sec:app_PubRad}

\input{Tables/Published_Radii}

\clearpage

\begin{figure*}
    \centering
    
  \includegraphics[width=23cm, height=14cm, angle=270 ]{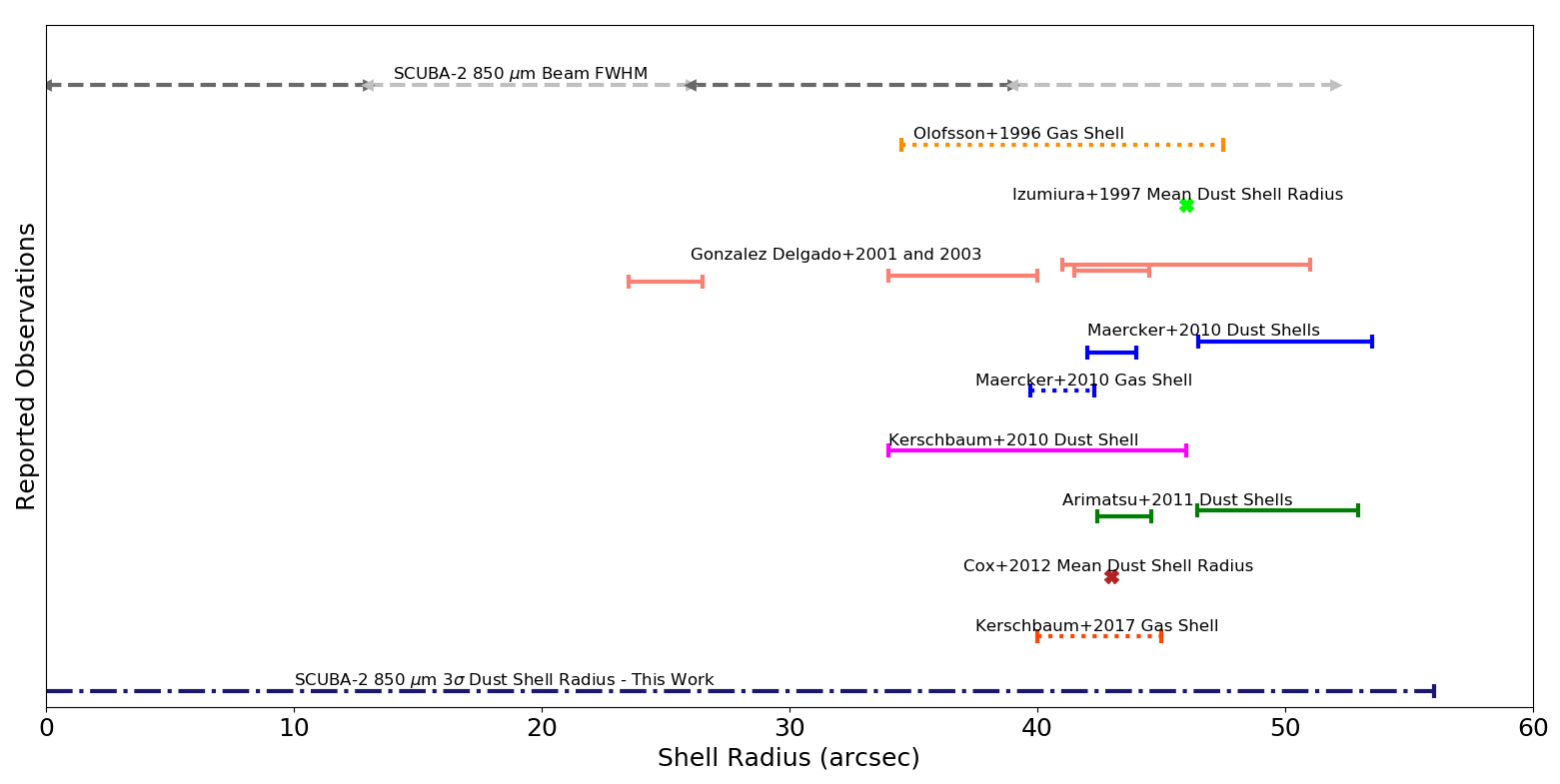}
  
\caption{Schematic diagram of the reported observations of U Ant showing mean Dust/Gas shell positions and FWHMs, arranged chronologically.  A summary of the observations are presented in Tab.~\ref{Table:PublishedRadii}. The top most grey dashed lines represent multiple SCUBA-2 $850~\micron$ beam FWHMs ($13\arcsec$).}
\label{fig:Schematic}
\end{figure*}

\clearpage



\section{CO-subtraction}
\label{Sec:app_HARPred}

\begin{figure}
   \centering
    \includegraphics[width=0.4\textwidth]{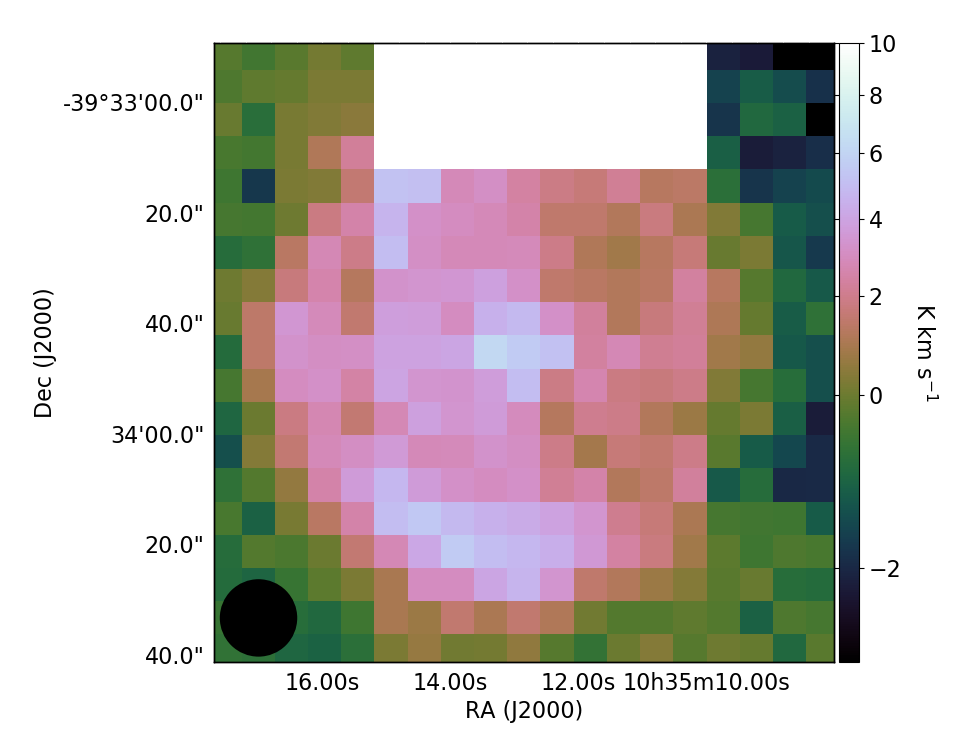}
    \caption{Integrated $^{12}$CO (3-2) HARP observation of U Ant used to carry out CO subtraction on the SCUBA-2 $850~\micron$ observation. Filled black circle in the bottom left corner: HARP beam with FWHM of $14\arcsec$. The figure is integrated over the velocity range of $\left[-26, 82\right]$ km s$^{-1}$}
  \label{fig:UAnt_HARP}
\end{figure}

In order to carry out $^{12}$CO(3-2) subtraction we use JCMT/HARP $^{12}$CO(3-2) heterodyne spectral observations obtained from 2017/01/03 to 2017/01/13, once again as part of the NESS survey. These maps use a 5$\times$5 jiggle pattern to produce a $2^\prime \times 2^\prime$ map, oversampled with $4.8^{\prime\prime} $ pixels. The HARP observations were reduced using standard JCMT Heterodyne \textsc{REDUCE\_SCIENCE\_NARROWLINE} pipeline \citep{Jenness_15_ORAC} and binned to 4 km/s resolution. See figure~\ref{fig:UAnt_HARP}. Using instructions provided by \citet{Parsons2018_SCUBA2_COsubtraction} (and in \textit{SCUBA-2 Data Reduction – Tutorial 5} webpage\footnote{\url{https://www.eaobservatory.org/jcmt/science/reductionanalysis-tutorials/scuba-2-dr-tutorial-5/}}) we generated the $^{12}$CO(3-2) subtracted SCUBA-2 $850~\micron$ observation. 

As seen in Fig~\ref{fig:UAnt_HARP}, HARP has two dead receptors meaning that no data was recorded for this section of the shell. Comparing the HARP observation to the ALMA CO observations by \citet{Kerschbaum2017} only the very edge of a small section of the shell falls within this missing pixel region and therefore has little effect on the CO flux ($\sim 15$ pixels out of 154 pixels within the shell). In addition, the chop throw was set to $60\arcsec$, smaller than the diameter of the shell, resulting in some self-subtraction. Between these two effects we estimate that approximately $30\%$ of the CO flux is missing and therefore incorporate additional uncertainty to account for this.

\clearpage

\section{Example of MCMC model fit}
\label{Sec:app_MCMC_Fitquality}

\begin{figure*}
   \centering
  \includegraphics[width=0.8\textwidth]{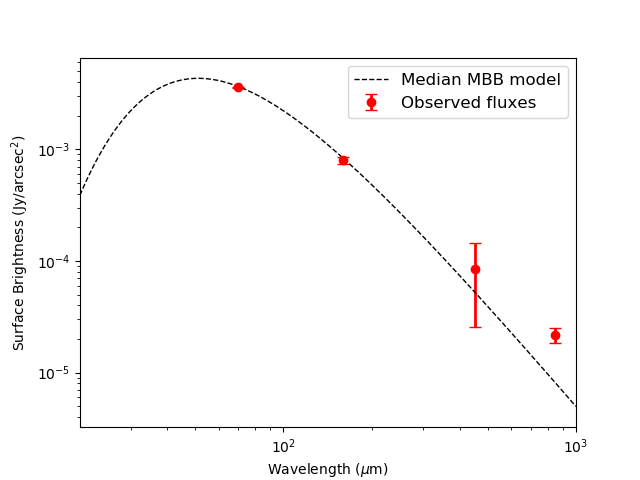}
   \caption{Residual profile surface brightness (red points) overlaid with a model (black line) whose parameters are equal to the median of the posterior samples generated by MCMC. The chosen radial point is at $40\arcsec$ (point 11).}
   \label{fig:MCMC_FitQuality}
\end{figure*}

The methods used by MCMC provides a representative value of the fit (in this case the median) of each parameter. There is no best fit model for the data when using MCMC methods. In Fig.~\ref{fig:MCMC_FitQuality} we have shown the modified black body model resulting from the median output values (from a set of $\sim 900000$ samples from the posterior) of the MCMC fit at the $40\arcsec$ residual profile radial point (point 11). This median is used as the parameter values presented in Fig.~\ref{fig:UAnt_T_B_D_profile}.

\clearpage
\section{Observed global SED fluxes of U Ant}

\input{Tables/UAnt_SED_table}

\clearpage

\section{SCUBA observations from 1997}
\label{sec:App:SCUBAobs}

U Ant was observed for project M96BI17 on 1997/10/17 and 1997/10/20 for a total of 2.1 hours. The data were re-processed using the SURF package, using the standard calibration factor for the 850N filter at the largest available aperture size of 60$\arcsec$ \citet{Jenness2002}. The reduction process included correction for opacity using skydips taken around the observations (yielding $\tau\left(850\micron\right)$ of 0.28-0.41 at zenith); cleaning with a 5-sigma clip, despiking, sky removal and bolometer weighting; and map reconstruction with median-regridding in 3 arcsec pixels, matching the native sampling of jiggle observations. The map was smoothed with a 9$\arcsec$ Gaussian to an effective resolution of approximately 17$\arcsec$ FWHM. There is no information in the map on scales larger than the 2$\arcmin$ chop throw, so the true zero level is poorly established. The surface-brightness profile is around an estimated overall flux-centroid of 10:35:13.0, -39:33:52 (J2000), south of the expected position of the star (attributed to a poor pointing model at far-south declinations). The noise is estimated from the dispersion among pixels in each annulus, converted to a standard error based on the number of independent beams within the annulus.


\begin{figure*}
\centering
\begin{subfigure}{0.45\textwidth}
  \centering
  \includegraphics[width=1\textwidth]{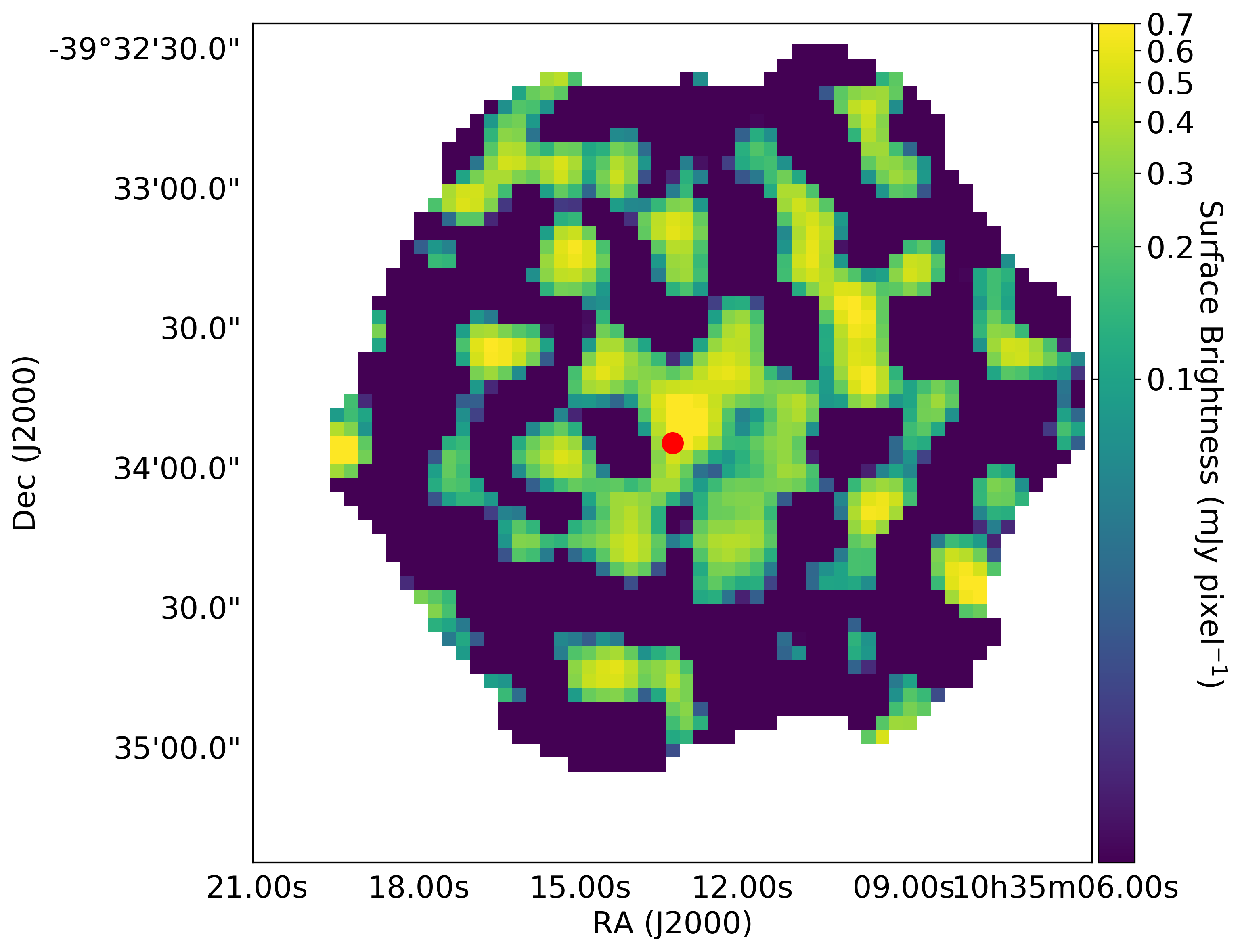}
  \subcaption{SCUBA $850~\micron$ observation}
  \label{fig:SCUBA_Obs}
  \end{subfigure}
\begin{subfigure}{0.45\textwidth}
  \centering
  \includegraphics[width=1\textwidth]{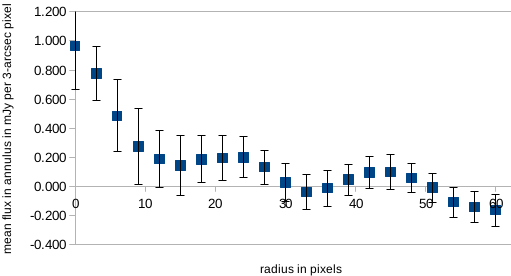}
  \subcaption{Surface-brightness profile}
  \label{fig:SCUBA_RadProf}
  \end{subfigure}
    \caption{(a): SCUBA $850~\micron$ observation of U Ant from 1997 (1 pix = $3\arcsec$). The off-centred red dot indicates the position of the star. It is off centre due to pointing accuracy problems in SCUBA; (b): Surface-brightness profile of the SCUBA observation.}
  \label{fig:UAnt_SCUBA}
\end{figure*}

%% file: Tables/Published_Radii.tex
\begin{table*}
  \centering
  \caption{Published Shell Radii of U Ant.}
    \begin{tabular}{llll}
    
    \hline
    \hline
    
    Publication & Mean  & Shell & Observation and \\
    & Shell Radius ($\arcsec$) & Thickness ($\arcsec$) & Shell Type \\
    \hline
    
    \citet{Olofsson1996} & 41 & 13 & SEST CO (1-0), (2-1), (3-2) - gas  \\
    &&& \\
    
    \citet{Izumiura1997} & 46 & -- & Far-IR IRAS - dust      \\
          & 180   & --  &  " \\
    &&& \\
    
    \citet{Gonzalez2001}  & 25    & 3   & Optical scattered light - dust    \\
    and   & 37    & 6  &  "\\
    \citet{Gonzalez2003}  & 43    & 3   &   " \\
          & 46    & 10    &  "  \\
     &&& \\      
    \citet{Maercker2010} & 43    & 2  & Optical scattered light - dust    \\
          & 50    & 7   &   " \\
          & 41    & 2.6   & APEX CO (3-2) - gas \\
     &&& \\      
    \citet{Kerschbaum2010} PACS & 40    & 12  & far-IR Herschel/PACS - dust   \\
     &&& \\
     
    \citet{Cox2012} & 42    & -- & Far-IR Herschel/PACS - dust \\

      &&& \\     
    \citet{Kerschbaum2017} & 42.5  & 5  &  ALMA CO (1-0) and (2-1) - gas \\

 &&& \\
 \hline
  &&& \\
    $3\sigma$ surface brightness  & 56    &   $-$ & Sub-mm SCUBA-2 $850~\micron$ - dust    \\
    extent derived in this paper &       &  &      \\
      &&& \\
    \hline
    
    \end{tabular}%
  \label{Table:PublishedRadii}%
\end{table*}%

%% file: Tables/UAnt_SED_table.tex
\begin{table*}
  \centering
  \caption{Fluxes used to derive the wavelength dependent SED of U Ant.}
    \begin{tabular}{llll}
    
   \hline
   \hline
    
    Instrument/Survey &  Wavelength  & Flux & Reference \\
    
    & ($\micron$) & (Jy) &\\
    
    \hline

    
    Gaia & 0.505 &	$16.8 \pm 0.3$ & \citet{Gaia2018_Cat}\\
    
    & 0.623 &	$73.1 \pm 0.3$ & \\\vspace{0.2cm} 
    
    & 0.772	& $154\pm 2$ & \\

    
    2MASS & 1.24  & 591 $\pm$ 151   & \citet{Skrutskie2006} \\
    
     & 1.66  & 1190 $\pm$ 370 & \\\vspace{0.2cm} 
    
     & 2.16  & 1270 $\pm$ 490 &  \\

              
    COBE/DIRBE & 1.25  & 614 $\pm$ 36  & \citet{Smith2004_COBE/DIRBE} \\
    
          & 2.22  & 1040 $\pm$ 30 &  \\
          
          & 3.52  & 725 $\pm$ 25  & \\\vspace{0.2cm}
          
          & 4.89  & 237 $\pm$ 10   & \\
   
   WISE & 11.6	& $111 \pm	27$ & \citet{Cutri2012_WISE_W3andW4Cat} \\\vspace{0.2cm}
   
   & 22.1 &	$35.8 \pm	0.3$ & \\

    AKARI/IRC & 8.61  & 264 $\pm$ 15  & \citet{Ishihara2010, Doi2015} \\\vspace{0.2cm}
    
          & 18.4  & 61.5 $\pm$ 2.3  &  \\
          

    AKARI/FIS & 65    & 25.8  $\pm$ 5.3 & \citet{Arimatsu2011} \\
          & 90    & 20.1  $\pm$ 4.2   &  \\  \vspace{0.2cm}
          
          & 140   & 8.4   $\pm$ 3.1   &  \\
          
          
    IRAS/ISSA & 11.6  & 168 $\pm$ 7  & \citet{Neugebauer1984, Beichman1988} \\
    
     & 23.9  & 44.8 $\pm$ 1.8 &  \\
    
     & 61.8  & 27.1 $\pm$ 2.7 &  \\\vspace{0.2cm}
    
     & 102   & 21.1 $\pm$ 2.3 &  \\
          
    Herschel/PACS & 70  & 23.06 $\pm$ 0.03 & Observations from \citet{Groenewegen2011} \\\vspace{0.2cm}
    
          & 160   & 5.96 $\pm$ 0.02 & - fluxes derived via aperture photometry in this paper \\
          
        
    Herschel/SPIRE & 250   & 1.81 $\pm$ 0.26 &  Observations from \citet{Groenewegen2011}\\
    
          & 350   & 0.716 $\pm$ 0.172 & - fluxes derived via aperture photometry in this paper  \\\vspace{0.2cm}
          
          & 500   & 0.243 $\pm$ 0.104 &  \\
          
          
    JCMT/SCUBA-2 & 450   & 0.435 $\pm$ 0.070 & This paper \\
    
          & 850   & 0.199 $\pm$ 0.034 &  \\
    
    \hline
    \end{tabular}%
  \label{table:UAnt_SED_Table}%
\end{table*}